\documentclass[journal]{vgtc}                
\ifpdf
  \pdfoutput=1\relax                   
  \usepackage{graphicx}                
  \DeclareGraphicsExtensions{.pdf,.png,.jpg,.jpeg} 
\else
  \usepackage{graphicx}                
  \DeclareGraphicsExtensions{.eps}     
\fi%

\graphicspath{{figures/}{pictures/}{images/}{./}} 

\usepackage{microtype}                 
\PassOptionsToPackage{warn}{textcomp}  
\usepackage{textcomp}                  
\usepackage{mathptmx}                  
\usepackage{times}                     
\usepackage{wrapfig}
\usepackage{cite}                      
\usepackage{tabu}                      
\usepackage{booktabs}                  

\usepackage{color,soul}
\usepackage{amsmath}
\usepackage{outlines}

\newcommand{\rev}[1]{\textcolor{black}{#1}}

\onlineid{0}

\vgtccategory{theory/model}
\vgtcpapertype{please specify}

\title{
A Bayesian cognition approach for belief updating of \rev{correlation judgement} through uncertainty visualizations 
}

\author{Alireza Karduni, Douglas Markant, Ryan Wesslen, Wenwen Dou}
\affiliation{\scriptsize University of North Carolina at Charlotte}
 \authorfooter{
 \item
  All authors are with the University of North Carolina at Charlotte. E-mails: $akarduni$, $dmarkant$, $rwesslen$, $wdou1@uncc.edu.$
 }

\shortauthortitle{Karduni \MakeLowercase{\textit{et al.}}: Bayesian cognition for correlation judgement}

\abstract{Understanding correlation judgement is important to designing effective visualizations of bivariate data. Prior work on correlation perception has not considered how factors including prior beliefs and uncertainty representation impact such judgements. The present work focuses on the impact of uncertainty communication when judging bivariate visualizations. Specifically, we model how users update their beliefs about variable relationships after seeing a scatterplot with and without uncertainty representation. To model and evaluate the belief updating, we present three studies. Study 1 focuses on a proposed ''Line + Cone'' visual elicitation method for capturing users' beliefs in an accurate and intuitive fashion. The findings reveal that our proposed method of belief solicitation reduces complexity and accurately captures the users' uncertainty about a range of bivariate relationships. Study 2 leverages the ``Line + Cone'' elicitation method to measure belief updating on the relationship between different sets of variables when seeing correlation visualization with and without uncertainty representation. We compare changes in users beliefs to the predictions of Bayesian cognitive models which provide normative benchmarks for how users should update their prior beliefs about a relationship in light of observed data. The findings from Study 2 revealed that one of the visualization conditions with uncertainty communication led to users being slightly more confident about their judgement compared to visualization without uncertainty information. Study 3 builds on findings from Study 2 and explores differences in belief update when the bivariate visualization is congruent or incongruent with users' prior belief. Our results highlight the effects of incorporating uncertainty representation, and the potential of measuring belief updating on correlation judgement with Bayesian cognitive models. 
} 


\CCScatlist{ 
 \CCScat{K.6.1}{Management of Computing and Information Systems}%
{Project and People Management}{Life Cycle};
 \CCScat{K.7.m}{The Computing Profession}{Miscellaneous}{Ethics}
}

\vgtcinsertpkg

\begin{document}

\maketitle

\section{Introduction}
Correlation judgement is an important topic and has been recently studied by the data visualization community \cite{yang2018correlation, harrison2014ranking, kay2015beyond, rensink_perception_2010}. 
Understanding how people perceive correlations from data is necessary for the design of effective visualizations like scatterplots.
Visualization researchers have investigated perceptual constraints on correlation judgment, including the use of Weber's Law \cite{harrison2014ranking, rensink_perception_2010}, a log-linear model augmented with censored regression and Bayesian methods \cite{kay2015beyond}, and other visual features \cite{yang2018correlation}.   
While these empirical studies and models provide valuable insights and recommendations for correlation visualization design, they can be expanded to consider other factors that affect people's understanding of variable relationships. 

One such factor is a user's prior beliefs when interpreting a correlation visualization. 
Previous studies often examine the perception of correlations between unnamed variables to avoid the effects of prior knowledge \cite{rensink2017nature,rensink_perception_2010} 
so that participants' beliefs about the variables do not influence their judgements.
However, in practice, people rely on prior knowledge when interpreting and learning from correlation visualizations.
As a result, it is important to investigate how prior beliefs affect the perception and interpretation of correlations.
In addition to prior beliefs, another factor related to correlation judgement that warrants more research is uncertainty communication. 
Recently, visualization researchers have argued for the importance of uncertainty communication in information visualization \cite{hullman2019authors}. 
Uncertainty communication techniques like hypothetical outcome plots (HOPs) \cite{hullman2015hypothetical,kale2018hypothetical} provide methods to visualize uncertain data for general audiences. 

The experiments in this paper build on previous research on correlation judgement by examining the impact of prior beliefs and uncertainty communication.
We explore the following research questions: (1) how do prior beliefs impact one's correlation judgement? (2) how do people adjust their beliefs when the correlation visualization aligns or conflicts with their prior belief? (3) when uncertainty communication is incorporated in a correlation visualization, are users more or less likely to adjust their beliefs based on the conveyed relationship? 


We also use Bayesian cognitive modeling \cite{griffiths2008bayesian} to quantitatively model how people interpret newly observed data in light of existing prior knowledge.
Bayesian cognitive modeling offers a principled framework to understand how people interpret visualizations in light of prior beliefs \cite{kim2017explaining} and how such beliefs should be updated with new information from a data visualization through Bayesian reasoning \cite{micallef2012assessing,ottley2015improving,kim2019bayesian}. 
This provides a normative framework 
for evaluating the effects of visualization on beliefs, including the impact of uncertainty communication on users' interpretations of data \cite{hullman2018pursuit,fernandes2018uncertainty,kim2017explaining}
and the presence of biases that impair data-driven decision making \cite{kim2019bayesian}.



Building such Bayesian cognitive models requires an accurate understanding of people's prior beliefs. 
Existing techniques for eliciting priors about correlations have a number of limitations, including a reliance on expert statistical knowledge related to correlation coefficients and their relationship to data \cite{kraan2002probabilistic,o2006uncertain,zondervan2017application}.
Our paper first evaluates a novel graphical elicitation method, ``Line + Cone'', for eliciting beliefs about the correlation between two variables through interactive data visualizations.  
With the proposed elicitation method, we conducted two experiments to study how people update beliefs about bivariate relationships when seeing correlation visualization with and without uncertainty representation.





\rev{This paper bridges several areas of past work on correlation judgment, belief elicitation, and uncertainty visualization, while also drawing on recent methods for modeling belief change using the framework of Bayesian inference.}
\rev{Specifically}, this paper's contributions are:
\begin{itemize}
    \item Study 1: Introduce and validate the \rev{graphical} ``Line + Cone'' method for eliciting prior beliefs about bivariate correlations, which is then used in the subsequent studies to measure belief change.
    \item Study 2: Compare differences in belief updating across correlation visualization with and without uncertainty communication. 
    \item Study 3: Explore differences in users' belief update when the correlation visualization (with and without uncertainty communication) is congruent or incongruent with their prior beliefs.
\end{itemize}

Analysis of Study 1 showed that the ``Line + Cone'' belief elicitation method \rev{can be used to estimate} peoples' mental representations of the correlation compared to a recent, \rev{more labor-intensive approach} from cognitive science \rev{for measuring subjective belief distributions} \cite{sanborn2008markov}. 
\rev{Study 2 revealed that participants updated their beliefs more effectively, and felt more confident, after observing visualizations with representations of uncertainty.}
In Study 3 we found evidence to support the hypothesis that people exhibit less belief change when seeing correlation visualizations \rev{that are incongruent with their prior beliefs.}
\rev{These results lay the groundwork for quantitative theories of how visualizations guide, and in some cases distort, how people learn about correlations through data visualization.}

\section{Background}


\subsection{Correlation perception and the effects of prior beliefs}


A common task in visual analytics is assessing the relationship between two or more variables, often as a scatterplot \cite{sedlmair2013empirical}. In statistics, such relationships are typically quantified as correlations. However, statistics like Pearson correlation can be misleading. For example, Anscombe's quartet \cite{anscombe1973graphs} demonstrates that hidden patterns in the data are obscured by identical statistics. Even for expert data analysts, visual data inspection is an important part of the analysis process. Past psychology studies have considered how perceptual processing of scatterplots can affect an individual's understanding of correlations \cite{meyer1997correlation,rensink_perception_2010,rensink2017nature}. Building off that research, InfoVis researchers have identified scatterplots as an effective technique in discriminating correlations \cite{li2010judging}, testing correlation perception with Weber's law through additional techniques \cite{harrison2014ranking,kay2015beyond}, and identifying visual features in correlation perception \cite{yang2018correlation}. However, these studies have not considered how prior beliefs affect individual's perception of variable relationships.

Research in psychology shows that prior beliefs have a strong influence on people's interpretation of uncertain data \cite{evans2005rapid,klaczynski2000motivated,sa1999domain,chinn1993role}, especially for correlations \cite{billman1992effects,baumgartner1995utility}. 
A central theory that explains why prior beliefs are important is the dual-process account of reasoning \cite{evans2003two,kahneman2011thinking}. 
This theory posits that fast heuristic processes (System 1) competes with slower analytic processes (System 2) that can affect logical decisions. Evans \textit{et al.} \cite{evans1983conflict} suggested that belief bias \cite{evans2003two,evans2005rapid} could occur as ``within-participant conflict'' between the two systems when participants tend to agree with an argument based on whether or not they agree with the conclusion rather than its logical conclusion. Alternatively, other research focused on theory-motivated reasoning bias based on  ``congruent'' and ``incongruent'' evidence relative to an individuals' belief systems \cite{klaczynski2000motivated}. These theories motivate design aspects in Study 2 and 3. 






\subsection{Uncertainty visualizations}

Uncertainty visualizations are important as they enable better decision-making by conveying the possibility that a point estimate may vary \cite{hullman2018pursuit}. 
More recently, research in InfoVis has provided innovative techniques like Hypothetical Outcome Plots (HOPs) \cite{hullman2015hypothetical,kale2018hypothetical}, frequency based representations \cite{kay2016ish,fernandes2018uncertainty}, visual semiotics \cite{maceachren2012visual}, and design guidelines \cite{greis2017designing} for visualizing uncertainty. Alternatively, other visualization researchers have studied important application aspects of uncertainty visualizations including hurricane prediction through ensemble modeling \cite{ruginski2016non,liu2016uncertainty}, comparing users' prior beliefs congruence to social data \cite{kim2017explaining}, how uncertainty evaluation is prone to error \cite{hullman2016evaluating}, and its potential to improve one’s ability to make predictions about replications of future experiments \cite{hullman2017imagining}.



\subsubsection{Eliciting correlation beliefs}

Psychologists have used a variety of approaches to elicit beliefs about correlations. Initial research used two-step procedure to elicit participant's correlation belief \cite{jennings1982informal, billman1992effects}: (1) determine relationship direction (positive or negative) and (2) rate the strength of the relationship. Later methods expanded on this approach by including Likert Scales, Spearmans’s correlation, probability of concordance, and conditional quantile estimates \cite{clemen2000assessing,kraan2002probabilistic,o2006uncertain,garthwaite2005statistical,johnson2010methods}. However, there are several shortcomings with the previous approaches. 
Some methods only elicit beliefs about central tendency without capturing degree of uncertainty, while methods which do elicit uncertainty are labor-intensive  \cite{zondervan2017application}. 
Most methods rely on some background knowledge of statistics \cite{kraan2002probabilistic,o2006uncertain}, including how to interpret correlation coefficients, thus limiting their applicability to non-expert populations.






Cognitive scientists have developed a related technique for eliciting subjective belief distributions named Markov Chain Monte Carlo with People (MCMC-P; \cite{sanborn2008markov,sanborn2010uncovering}).
Inspired by algorithms for MCMC estimation \cite{garthwaite2005statistical}, MCMC-P as an approach to estimate a person's subjective belief distribution through sampling. 
In Study 1, we use MCMC-P as an elicitation benchmark to our proposed Line + Cone belief elicitation technique and outline this technique in Section 4.




\subsection{Bayesian cognitive modeling in data visualizations}

Cognitive modeling in visualization initially was studied as a subset of visuospatial reasoning in how individuals derive meaning from external visual representations \cite{tversky2005visuospatial}. 
Visualization researchers have integrated similar ideas to understand visualization cognitive processes through insight-based approaches \cite{green2009building} and top-down modeling \cite{liu2010mental,patterson2014human}. 
More recently, InfoVis researchers have used Bayesian models to understand cognitive processing of visualizations \cite{wu2017towards,kim2019bayesian}.
Cognitive scientists have demonstrated the importance of Bayesian modeling to understanding individual decision-making \cite{griffiths2006optimal,griffiths2008bayesian}. 
In this approach, an individual has some prior belief that is updated when the individual consumes additional data, resulting in their posterior beliefs. Bayesian cognition models have been used to understand deviations from optimal belief updating due to conservatism, sample-based inference (approximation) and ``resource-rational'' interpretations of cognitive bias \cite{lieder2018resource}. 

 To our knowledge only two previous InfoVis studies \cite{wu2017towards,kim2019bayesian} have combined belief elicitation with a Bayesian cognitive modeling framework. Wu \textit{et al.} \cite{wu2017towards} examined whether people integrated prior probabilities with data in an optimal manner. They found that priors influenced predictions in a manner consistent with Bayesian inference, although to a lesser extent than predicted by the model. However, a limitation to this study was that participants were given a prior; therefore, prior beliefs cannot be examined. In contrast, Kim \textit{et al.} \cite{kim2019bayesian} empirically measured participants' prior beliefs about the a target proportional quantity and used those priors to calculate the normative posterior given the data that was presented.
In aggregate, participants' judgments were consistent with predictions derived from Bayesian inference, although less so for large data sets. However, participants expressed greater uncertainty in their judgments than expected from the Bayesian model.
Further, the authors connect such Bayesian modeling and belief elicitation with recent research on visualizing uncertainty through techniques like HOPs \cite{hullman2015hypothetical,kale2018hypothetical}. Our work extends their framework but considering correlation beliefs rather than proportional values.

\section{Research Questions and Analysis Methods}

Our primary research question is the effect of providing uncertainty communications on users' belief updating in correlation visualization. In order to address this research question, we conducted a sequence of three experiments with latter ones building on the earlier studies. 

A key to understanding users' belief update is the ability to accurately and intuitively capture such beliefs. \textbf{Study 1} evaluates the Line + Cone elicitation method relative to Markov Chain Monte Carlo with People (MCMC-P) \cite{sanborn2008markov}, a belief elicitation method from cognitive science. 
After validating the Line + Cone method,
we apply it in the next two experiments to address the main research question. In \textbf{Study 2}, we explore the effect of correlation visualizations with and without uncertainty representation on belief updating. Our primary hypothesis is that visualizations with uncertainty representation will overall lead to less belief updating about the correlation between two variables. 
Findings from Study 2 provides partial evidence to support the primary hypothesis. To expand on the findings, we are interested in further understanding users' belief update when the data visualization was deliberately manipulated based on users' prior beliefs.
Therefore, \textbf{Study 3} extends Study 2's design but introduces a treatment that alters the data provided to participants to be either congruent or incongruent with their prior beliefs. 
We then evaluate the degree to which individuals update their beliefs when data provided either conflicts or aligns with their prior and whether the presence of uncertainty visualizations interact with that effect. 

To analyze the results of Study 2 and 3, we employ mixed effects models to identify differences between treatments. The mixed effects models control for individual heterogeneity assumed between participants and the datasets (variable pairs) provided to participants. To explain the findings from the mixed effects models, we evaluate whether Bayesian cognitive models can be used to predict users' posterior beliefs under different experiment treatment.


\section{Study 1: Evaluating Line + Cone Elicitation}

\begin{figure}[t] 
\centering%
\includegraphics[width=.48\textwidth]{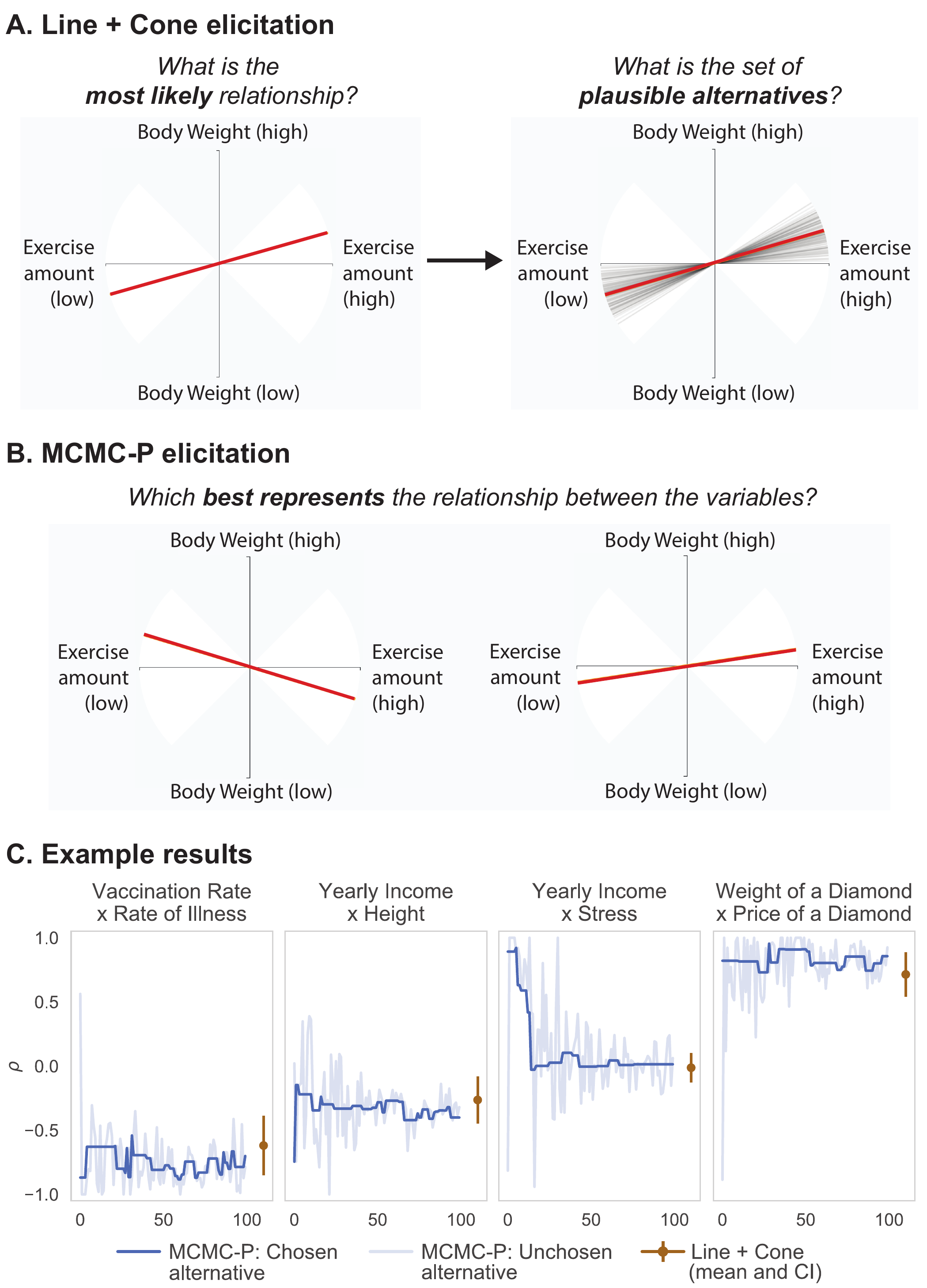}
\caption{Elicitation methods in Study 1. \textbf{A:} For the Line + Cone elicitation, participants first recorded the belief about the most likely relationship between two variables (red line), then adjusted the set of plausible alternatives based on their uncertainty (gray lines). \textbf{B:} For the MCMC-P elicitation, participants responded to a series of two-alternative forced choices in which they judged which of two lines was more likely to represent the true relationship between the variables. \textbf{C:} Example comparison of elicitation results for a participant in Study 1. Dark blue lines indicate the chain of chosen alternatives from MCMC-P across 100 trials. Light blue lines indicate unchosen alternatives. The corresponding mean and CI from the Line + Cone elicitation is shown at the right of each plot.}
\label{fig:study1_task}
\end{figure}


Our goal in Study 1 (see preregistration\footnote{http://aspredicted.org/blind.php?x=zp7hr3}) was to develop and validate the Line + Cone visual interface for eliciting prior beliefs about the correlation between two variables.
In selecting our approach, we aimed to 
measure beliefs about both the \emph{most likely} correlation between variables and the \emph{degree of uncertainty}, \rev{without a need for statistics domain knowledge or numerical reasoning (see Section 2.2.1).}
\rev{We assessed the convergent validity of the Line + Cone method by comparing it to a higher resolution, but more labor-intensive, approach to eliciting subjective beliefs:} Markov Chain Monte Carlo with People (MCMC-P; \cite{sanborn2008markov,sanborn2010uncovering}).
MCMC-P resembles common sampling-based estimation algorithms such as Metropolis-Hastings in which a chain of states are sampled from an underlying probability distribution.
In MCMC-P, state transitions are determined by asking participants to make forced-choice comparisons of the likelihood of possible values of the target parameter in many trials (usually in the range of 100 or more).

In our experiment we elicited prior beliefs about five sets of variables using both MCMC-P and the Line + Cone method. \rev{We created variable sets to cover a range of plausible correlations
possible divergent prior beliefs. For example, we expected that for the relationship \textit{Weight x Price of diamonds} most participants would believe there is a strong positive correlation, while there may be less consensus about the relationship \textit{Vaccination rate x Rate of illness}.} 
Based on participants' responses we estimated the mean and confidence interval of their subjective prior belief (i.e., the relative likelihood of possible correlations between two variables).
We then examined the degree to which the resulting prior means and CIs were correlated across the two methods.

\subsection{Study Design}

The experiment involved a within-subjects manipulation of elicitation method (Line + Cone vs. MCMC-P). 
Participants' beliefs were elicited for the same set of five variable sets (Table \ref{tab:study1_results}) using each method \rev{in a blocked presentation}.
The order of elicitation methods and variable sets within each block were randomized for each participant.


\subsubsection{Line + cone elicitation}

We designed a visual interface in which the mean and CI are directly elicited through the user's interaction. 
Each elicitation involves a two-step procedure (Figure \ref{fig:study1_task}A).
First, the user selects the orientation of a red line according to their belief about the most likely relationship between the variables.
Second, the user adjusts the width of the uncertainty cone. 
The uncertainty cone was depicted by gray lines which were draws from a Normal distribution centered on the most likely correlation (red line) and truncated at -1 and 1.
Participants were instructed to adjust the cone such that the lines captured the range of ``plausible alternatives" for the relationship between the variables.

\subsubsection{MCMC-P elicitation}

Markov Chain Monte Carlo with People (MCMC-P) is used to estimate subjective belief distributions based on a series of choices between two alternatives. 
In our task, each alternative represents a potential correlation between a pair of variables.
For each variable set there were 100 choice trials.
On each trial, the participant was shown two lines representing potential correlations (Figure \ref{fig:study1_task}B). Participants were instructed to select the alternative which was more likely to represent the true relationship.
On the first choice trial the alternatives were two randomly selected correlations, one positive and one negative.
In subsequent trials, the choice set included the alternative chosen on the previous trial and a \emph{proposal} generated from a Normal distribution centered on the previous choice.
The width of the proposal distribution was adaptively tuned based on how often a participant accepted new proposals (see \cite{roberts_optimal_2001}).
Each block resulted in a chain of alternatives that were chosen by the user (Figure \ref{fig:study1_task}C).
The prior mean was calculated as the mean of the sampling chain, while the CI was the range between the 2.5\% and 97.5\% quantiles.

\begin{figure*}[t] 
\includegraphics[width=0.8\textwidth]{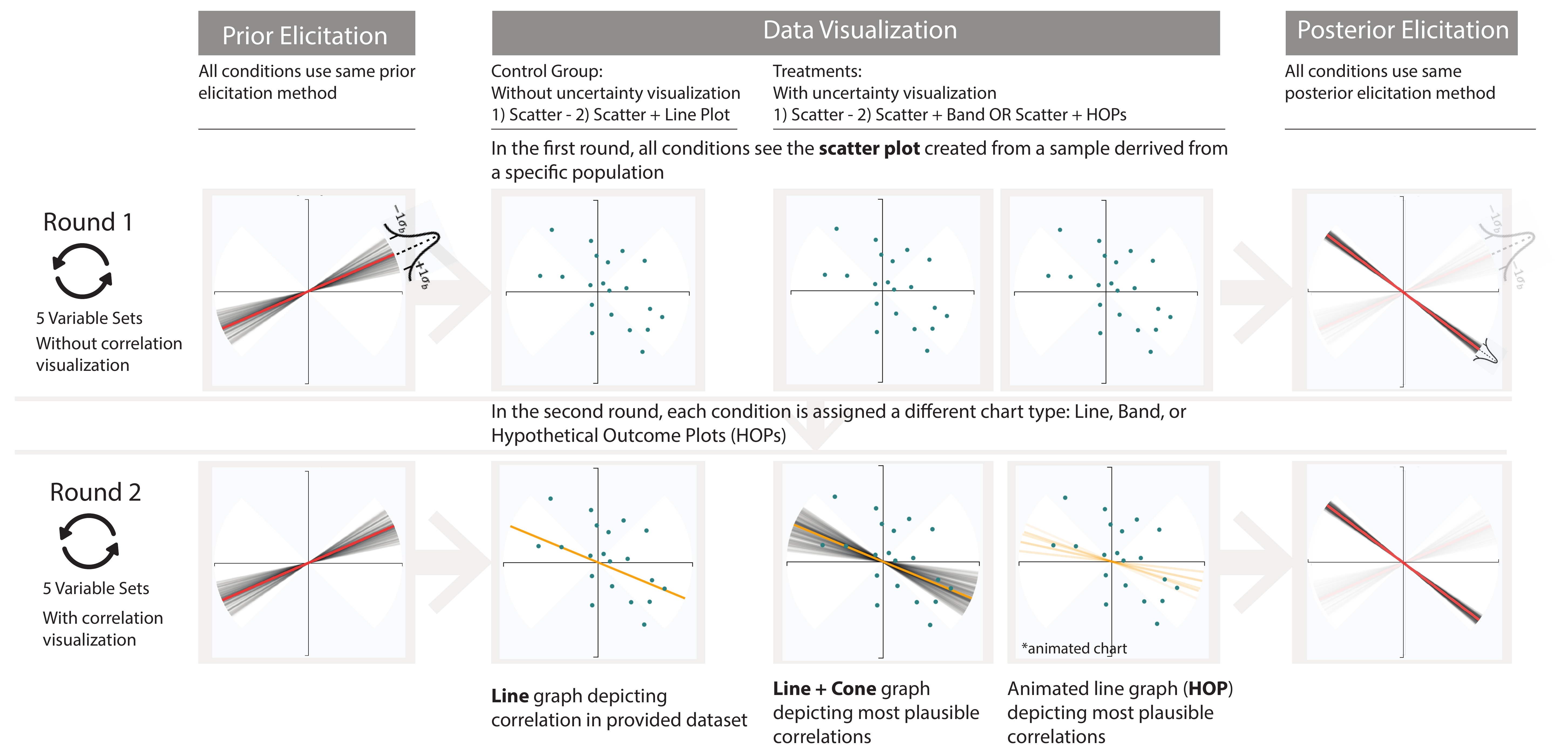}
\centering
\caption{Study 2 Design. Each user goes through ten variable sets (five variables for two rounds) and elicit their belief before and after seeing data visualizations about each variable set. In Round 1, the user views five variable sets through only scatterplots. In Round 2, the user is randomly assigned to either Line, Cone, or HOP visualization treatments and views the remaining five variable sets.}
\label{fig:study2}
\end{figure*}

\subsection{Participants}

$N = 152$ participants were recruited from Amazon Mechanical Turk.
Participants earned \$2.00 upon completion of the task, which took an average of 25.4 minutes ($SD =12.2$).
Per our pre-registration, we used several measures of task engagement to decide whether to exclude a participant.
We excluded 55 participants who failed an attention check question 
and 36 participants who made nonsensical or incomplete responses to a set of open-ended questions regarding how they would respond to real-world situations.
We also excluded $35$ participants who met pre-specified exclusion criteria based on responses in the MCMC-P elicitation, including response streaks, response alternation, and response time.
After accounting for all exclusions, $N = 92$ participants were included in the analysis.

\subsection{Results and Discussion of Study 1}
\begin{table}[h]
  \caption[width=1\columnwidth]{Correlations between prior means and CIs elicited through Line + Cone and MCMC-P methods in Study 1.}
  \label{tab:study1_results}
  \resizebox{\columnwidth}{!}{%
  \begin{tabular}{p{0.25\textwidth} p{0.07\textwidth} p{0.06\textwidth} p{0.07\textwidth} p{0.06\textwidth}}
  \toprule
  & \multicolumn{2}{c}{Prior mean} & \multicolumn{2}{c}{Prior CI} \\
  \cmidrule(lr){2-3}\cmidrule(lr){4-5}
  Variable set & Pearson $r$ & $p$-value & Pearson $r$ & $p$-value \\
  \midrule
  Weight x Price of diamonds          & .26 & \textbf{.012}     & .34 & \textbf{.001} \\
  Exercise amount x Body weight       & .37 & \textbf{$<$ .001} & .29 & \textbf{.005} \\
  Yearly income x Height              & .12 & .26               & .27 & \textbf{.010} \\
  Yearly income x Stress              & .45 & \textbf{$<$ .001} & .30 & \textbf{.003} \\
  Vaccination rate x Rate of illness  & .40 & \textbf{$<$ .001} & .39 & \textbf{$<$ .001} \\
  \bottomrule
  \end{tabular}
  }
\end{table}

\noindent Our primary question was whether the belief distributions elicited with the Line + Cone method correlated with those generated using our MCMC-P procedure.
We calculated Pearson correlations between the prior means and CIs for each variable set (Table \ref{tab:study1_results}).
Elicited prior means were significantly correlated for 4 of the 5 variable sets, with the \emph{Yearly income X Height} variable set the only exception.
Prior CIs elicited from the two methods were significantly correlated in all 5 variable sets.
These results suggest that our visual Line + Cone elicitation method is able to capture variation in beliefs about correlations across different variable sets, including beliefs about the most likely relationship as well as the degree of uncertainty, while being less labor-intensive than MCMC-P and \rev{requiring less statistics domain knowledge} than existing elicitation methods.

\section{Study 2: Belief updating with and without uncertainty representations}

In the second study, we applied the Line + Cone elicitation method to examine belief change in the context of correlation visualization.
We evaluated whether the type of visualization impacted the degree to which people updated their beliefs.
Specifically, our \textbf{Study 2 Main Hypothesis} \footnote{http://aspredicted.org/blind.php?x=39yn5g} was that correlation visualizations which include representations of the uncertainty in the true population correlation would lead to less belief updating when people's prior beliefs were inconsistent with the presented data. 
\rev{This hypothesis is motivated by research on confirmation bias \cite{mynatt1977confirmation,nickerson1998confirmation} showing that people overweight evidence that is consistent with their prior beliefs. 
Uncertainty visualizations, by giving credence to a range of possible relationships (including less likely relationships that are more similar to a person's prior belief) may lead to less belief updating compared to visualizations that only represent the most likely a posteriori relationship.}
As a secondary hypothesis, we hypothesize that datasets with small and moderate correlations lead to less belief updating compared to datasets with \rev{stronger} correlations. 

\subsection{Study Design}

We employed a mixed design with a between-subjects manipulation of the visualization type (with and without uncertainty representation) and a within-subjects manipulation of the sample correlation of data presented to participants.
In each trial participants reported their belief about the relationship between a set of variables, both before and after they experienced a data visualization 
All participants completed two rounds of five trials.
In the first round the datasets were visualized as 
scatterplots to all participants (\textbf{Scatter} condition). 
In the second round the scatterplots were augmented with a visualization of the predicted population correlation based on the given dataset. Participants were randomly assigned to one of the following 
conditions 
(Fig \ref{fig:study2}):

\begin{itemize}
    \item \textbf{Line}: A line representing the most likely population correlation was superimposed on the scatterplot \footnote{Note that the Line condition does not contain an uncertainty representation while the Cone and HOP conditions do.} 
    \item \textbf{Cone}: The line appeared with an uncertainty cone which represents the 95\% confidence interval for the population correlation
    \item \textbf{HOP}: Hypothetical outcome plots (HOPs, \cite{hullman2015hypothetical,kale2018hypothetical}) were used to present animated draws from the 95\% confidence interval for the correlation
\end{itemize}

\subsubsection{Datasets} 

We created two groups of five variable pairs that covered a range of population correlations between -0.9 to 0.9.
We then generated 100 random samples for each variable pair based on the population correlation. \rev{The participants were told that the dataset is a \textit{a sample of data collected from the real world.}}
\footnote{Due to random sampling, the sample correlations differed slightly from the specified population correlation.}
All points were re-centered with a mean of zero on each variable.
All participants saw the same data points for each variable pair.
The order of the variable pairs was randomized for each participant.

Note that population correlations were specified for each variable pair based on agreement among the authors (see examples in Figure \ref{fig:study2_beliefs}).
\rev{Our assumptions about the correlations of these variables may not reflect the ground truth relationship, and may differ from participants' beliefs. 
However, because we measure each individual's prior beliefs, we can assess whether belief updating was affected by any mismatch between their prior and the sample correlation.}

\subsubsection{\rev{Elicitation, attention check procedures, and collected data}}

Each trial consisted of a prior elicitation, correlation visualization, and posterior elicitation. 
For both elicitation steps we used the Line + Cone method validated in Study 1 (Figure \ref{fig:study1_task}A).
Each elicitation resulted in three measurements: the most likely correlation ($\mu$) and the lower and upper bounds of the uncertainty cone ($b_{lower}, b_{upper}$). 
All three values were bounded between $\rho = -1$ and $\rho = +1$.



We designed practice questions to familiarize participants with the Line + Cone elicitation.
Participants answered test questions to ensure that they understood how to interpret the elicitation interface, including the direction of a correlation and the degree of uncertainty captured with 
the cone.
We also included attention check questions (same as in Study 1) to screen inattentive respondents or other invalid data~\cite{chmielewski2019mturk}.

\rev{In Study 2 and 3, we also collected basic demographic data, duration of each trial, and the error count of users in the instructions section.}

\subsection{Participants}

Participants were recruited from Amazon Mechanical Turk.
For all studies we required that participants were located in the U.S. and had a 95\% or above approval rating.
Participants earned \$1.80 upon completion of the task, which took an average of 25.7 minutes ($SD = 14.6$) to complete.
Per our pre-registration, we excluded any participants due to: failed attention check questions ($n=35$); technical errors ($n=15$); or task completion in less than 5 minutes ($n=38$). 
This left $n=212$ participants for the analysis (Line: 74; Cone: 64; HOP: 74).

\subsection{Results}

For the analysis we built three mixed effects models using R's \texttt{lme4} package for two linear regressions and R's \texttt{glmmTMB} for a beta regression. We used the normal
approximation to calculate p-values of fixed effects using t-values produced by \texttt{lme4}.\footnote{The code used is included in our supplemental materials.}

\textbf{Dependent \& Independent Variables:} We considered three dependent variables (DV): (1) the absolute belief difference, (2) the difference in uncertainty, and (3) belief distance from the model's predicted posterior mean.  For our independent variables (IV), we included the Visualization treatment (Line, Cone, HOP, and Scatter) and the absolute correlation of the generated data for the variable sets 
(see Figure 3)

\textbf{Model Specification:}
For each model, we included the visualization treatment and the absolute correlation of the data as fixed effects. For the visualization treatment, the Scatter condition is the omitted reference condition. We treated the sample correlation as a categorical variable and used zero absolute correlation as the omitted reference condition. We included the unique variable set and the participant id as random effects.

\subsubsection{\rev{Beliefs about variable pairs}}
\begin{figure}[h] 
\includegraphics[width=1\columnwidth]{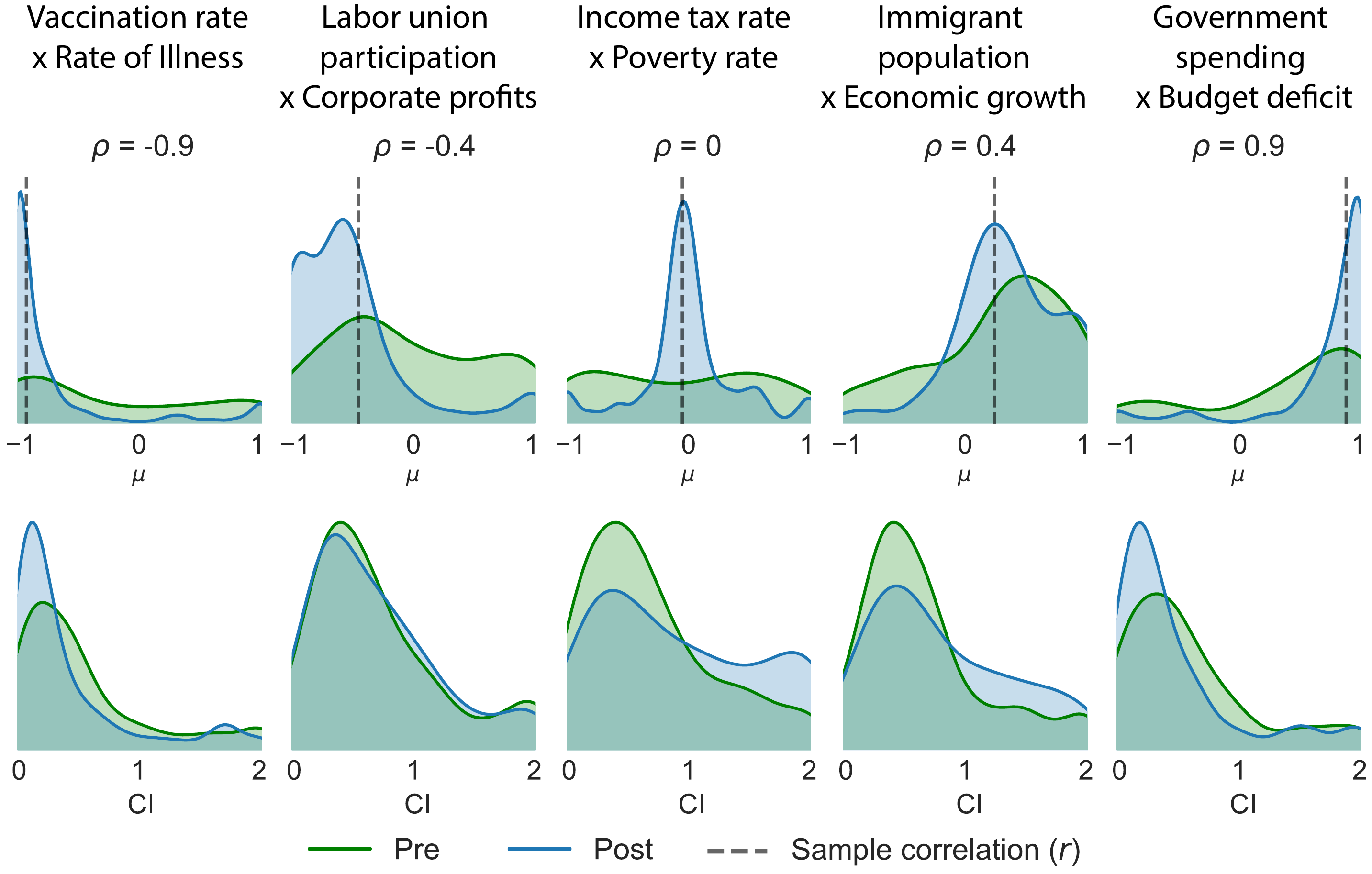}
\centering
\caption{Density plots of means (top row) and CIs (bottom row) of elicited belief distributions for selected variable sets in Study 2. Dashed lines indicate the sample correlation of the dataset presented to participants.}
\label{fig:study2_beliefs}
\end{figure}

We first examined participants' beliefs  before and after experiencing the data visualization.
Figure~\ref{fig:study2_beliefs} displays pre- and post-treatment judgments about the most likely correlation ($\mu$, top row) and uncertainty ($CI$, bottom row) for five of the ten variable pairs, aggregated across visualization treatments.
With respect to the mean correlation $\mu$, prior judgments (green density plots) were largely consistent with the relationship that was designated for each variable pair, such that the modal prior belief was close to the sample correlation.
This suggests that the datasets presented were congruent with most participants' prior belief about the relationship between the variables.
One notable exception was \textit{Income tax rate X Poverty rate}, where the designated correlation was $\rho = 0$ but prior beliefs were relatively uniformly distributed from -1 to +1.
Post-treatment beliefs about the same variable sets (blue density plots) strongly shifted toward the sample correlation of the observed dataset (dashed lines) for all variable sets.
\rev{The plots for the CIs reveal that the strength of the sample correlation also affected changes in uncertainty. CIs decreased after seeing strongly correlated datasets ($\rho = \pm0.9$) but in some cases increased following data visualizations with weaker relationships.}
We report more detailed analysis of how the uncertainty changed in different treatments in section 5.3.3. 





\begin{figure}[b] 
\includegraphics[width=1\columnwidth]{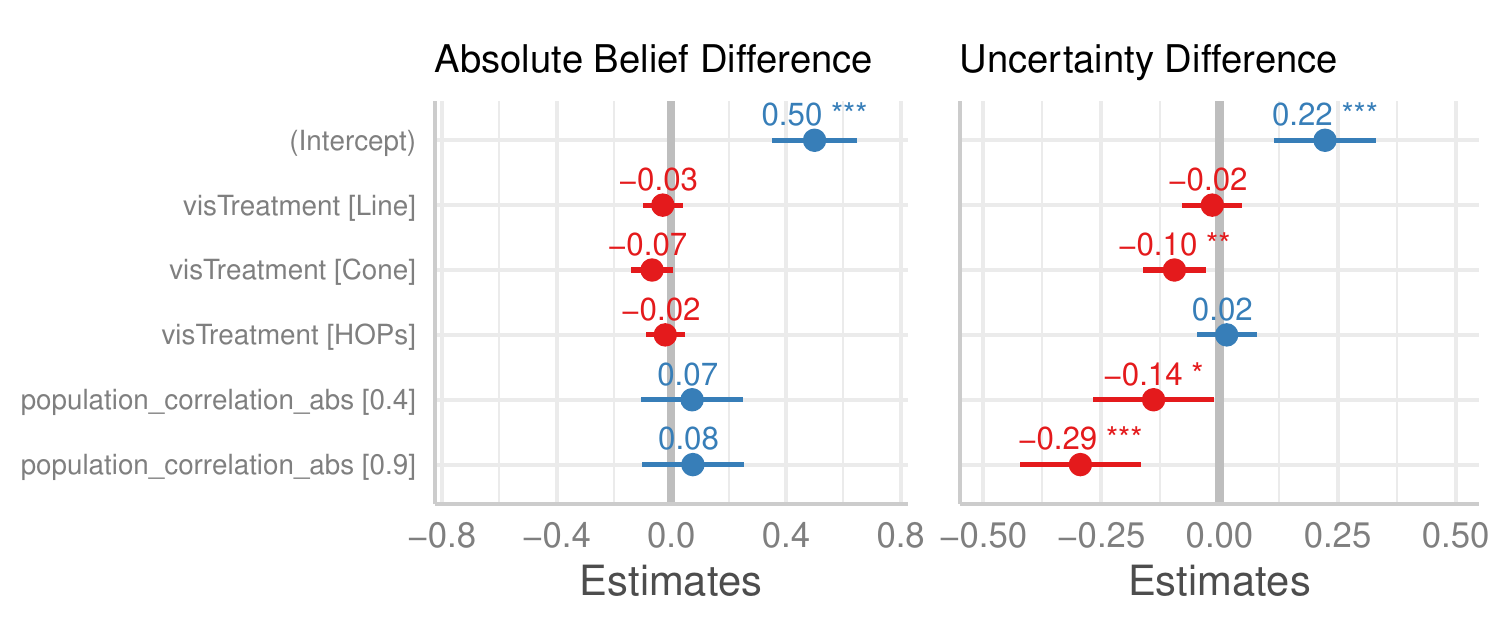}
\caption{Study 2 fixed effects coefficients for absolute belief difference (left) and uncertainty difference (right). Error bars indicate 95\% confidence intervals. 
Asterisks indicate statistical significance than zero using p-values: *** 99.9\%, ** 99\%, * 95\%. For visTreatment, the reference category is the Scatter condition.}
\label{fig:study2_me}
\end{figure}

\subsubsection{Change in beliefs about most likely relationship}


We used linear mixed effects regression to model the effect of visualization conditions and population correlation on the absolute change in beliefs about the most likely correlation ($|\mu_{post} - \mu_{pre}|$.
There were no significant effects (Figure \ref{fig:study2_me}, left), though
the Cone condition showed marginally smaller changes in beliefs
compared to the Scatter condition $(\beta = -0.07 \;\;  [-0.14, 0.01], \;\; z = -1.779, \;\; p = 0.075)$. 
Thus, while participants clearly shifted their beliefs about the most likely correlation in response to observed datasets (Figure Figure~\ref{fig:study2_beliefs}), contrary to our expectations we did not find that the degree of belief change differed by visualization treatment or population correlation.



\subsubsection{Change in uncertainty}

Mixed effects linear regression was used to model the effects of visualization condition and population correlation on the change in uncertainty ($|CI_{post} - CI_{pre}|$). 
As shown in Figure \ref{fig:study2_me} right, relative to the Scatter condition, the Cone condition exhibited greater reduction in uncertainty $(\beta = -0.10 \; [-0.16, -0.03], \; z = -2.782, \; p < .01)$. \rev{In other words, participants assigned to the cone Condition felt less uncertain (more confident) with their input.}
There was no difference in the Line $(\beta = -0.02 \;  [-0.08, 0.05], \; z = -0.468, \; p = 0.640)$ or HOP condition $(\beta = 0.02 \;  [-0.05, 0.08], \; z =  0.473, \; p = 0.636)$.
In addition, more extreme sample correlations had a greater impact on belief change:
Compared to $\rho = 0.0$, there was a greater reduction in uncertainty for $\rho = .4$ ($(\beta = -0.14 \;  [-0.27, -0.01], \; z = -2.138, \; p < .05)$ and $\rho = .9$  ($\beta = -0.29 \;  [-0.42, -0.17], \; z = -4.513, \; p < .001)$ variable sets. 

\subsubsection{Accuracy of posterior beliefs}

We examined the accuracy of participants' posterior mean ($\mu_{post}$) compared to the sample correlation of the observed datasets.
In the Scatter condition, posterior means were biased to be more extreme for moderately positive and negative sample correlations.
Relative to the $\rho = 0$ variable sets, absolute error was higher for $\rho=\pm0.4$ variable sets ($\beta = .29 \; [.19, .41], z = 5.46, p < .001$) but did not differ from $\rho=\pm0.9$ variable sets ($\beta = .002 \; [-.11, .11], z = .03, p = .96$)
The remaining visualization conditions led to more accurate beliefs across the full range of sample correlations.
Compared to the Scatter condition, the absolute error was lower in all three visualization conditions (Line: $\beta = -.22 \; [-.34, -.10], z = -3.50, p < .001$; Cone: $\beta = -.34 \; [-.47, -.21], z = -4.96, p < .001$; HOP: $\beta = -.25 \; [-.38, -.13], z = -3.97, p < .001$).

\subsection{Bayesian belief updating model}

\setlength{\intextsep}{-6pt}%
\setlength{\columnsep}{4pt}%
\begin{wrapfigure}{L}{0.26\textwidth}
  \vspace{0pt}
  \begin{center}
    \includegraphics[width=0.23\textwidth]{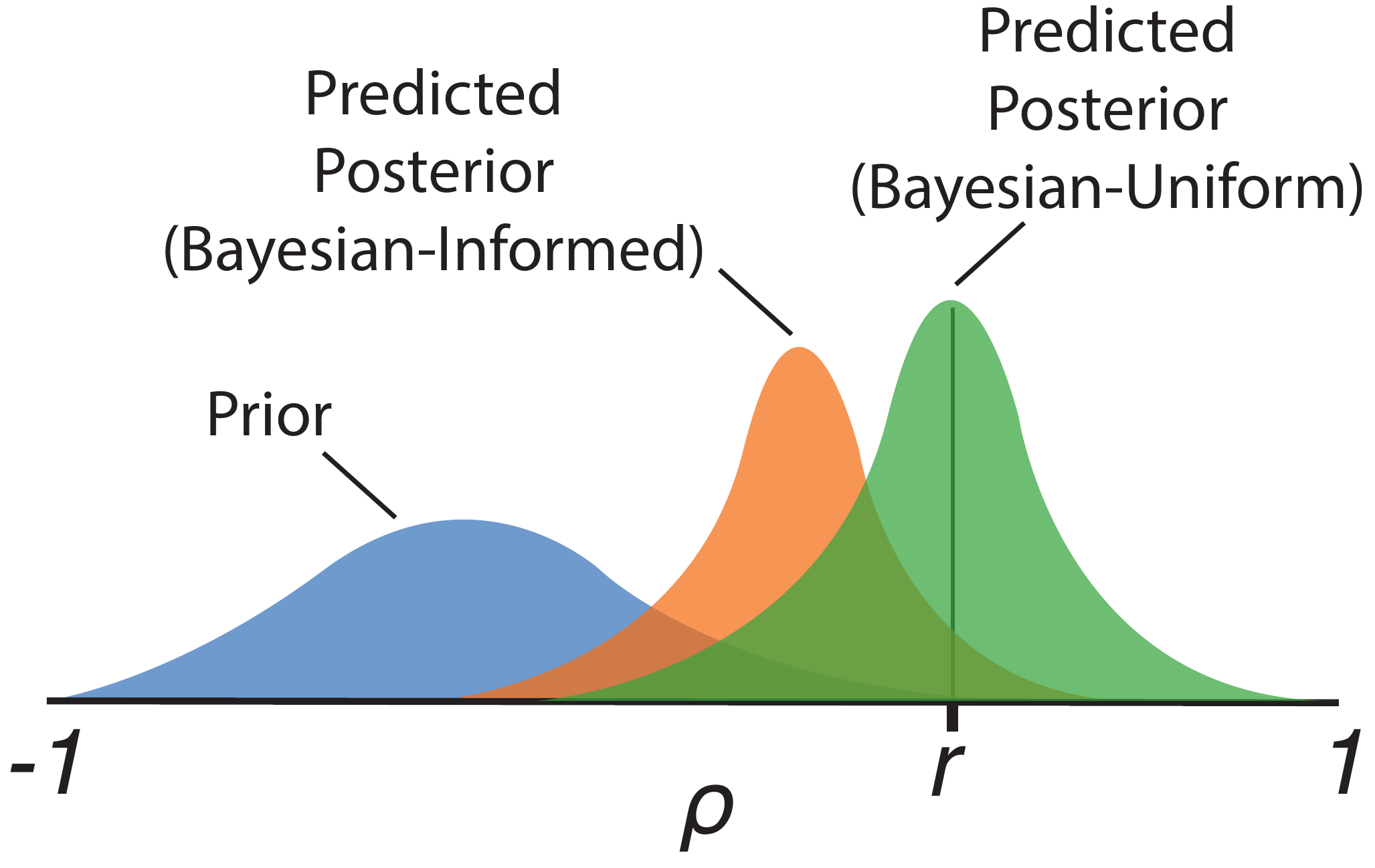}
  \end{center}
\end{wrapfigure}


In this section we use Bayesian cognitive modeling to investigate the influence of prior beliefs on the belief updating process.
Under the principles of Bayesian inference, people should integrate new evidence about a correlation with their prior beliefs about that relationship. 
Bayesian models provide a normative benchmark for how beliefs \emph{should} change depending on the strength of the evidence and participants' uncertainty.
For instance, a person who is confident that variables are negatively correlated may only shift their beliefs a small amount after seeing a dataset with a positive sample correlation.
A second person who is highly uncertain about the relationship, however, may be more strongly influenced by the same data and report posterior beliefs that are closely matched to the sample correlation.
\rev{This framework also allows us to identify when people systematically fail to adjust their beliefs as predicted by the Bayesian model.
Returning to the main hypothesis of Study 2, if uncertainty representations cause smaller adjustments to beliefs, this will correspond to larger divergence between participants' elicited posterior beliefs and the predictions of the Bayesian model compared to other conditions.}

Having elicited prior beliefs about each set of variables, we examined whether participants' posterior beliefs (following the data visualization) could be predicted by a normative Bayesian model.
The model uses Bayesian inference to predict a posterior belief distributions over possible population correlations, $\rho$, based on an observed dataset and a particular prior (see \cite{lee2014bayesian} for similar model formulation). 
We evaluated two variants of the model that differed only in their prior. 
The \textbf{Bayesian-Informed} model relied on the participant's elicited prior to calculate the normative posterior distribution after observing a dataset.
The prior belief was modeled as a bounded Normal distribution, $\rho \sim BoundedNormal(\mu_{pre}, \sigma_{pre}, [-1, 1])$, where $\mu_{pre}$ and $\sigma_{pre}$ are the mean and standard deviation of the participant's elicited prior.
The observed bivariate data $X$ was modeled as having been generated from a standardized multivariate Normal distribution with mean of zero and standard deviation of 1 on each dimension (see \cite{lee2014bayesian}),

\begin{equation}
    X \sim MultivariateNormal(\begin{bmatrix} 0 & 0 \end{bmatrix}, \begin{bmatrix} 1 & \rho \\ \rho & 1 \end{bmatrix}^{-1}).
\end{equation}

\noindent Under the \textbf{Bayesian-Uniform} model, the prior was a uniform distribution over the correlation coefficient, $\rho \sim Uniform(-1, 1)$.
The mean and 95\% CI of the posterior distribution for this model is equivalent to the values used for the visualizations in the Line, Cone, and HOP treatments. 
Predicted posterior distributions for $\rho$ were estimated for both models using MCMC with the PyMC3 library \cite{salvatier2016probabilistic} with two chains of 20,000 samples and 1000 burn-in iterations. 
Lastly, we compared the elicited posteriors to the elicited priors, absent any belief updating.
We refer to this baseline as the \textbf{Prior-only} model in the results below.

\rev{The relative fit of the models reflects the weight of prior beliefs in the updating process, with the Bayesian-Informed model representing the normative integration of priors with new evidence.} If people relied only on the visualization without accounting for prior beliefs, their elicited posteriors should be best fit by the Bayesian-Uniform model.
In contrast, if they did not adjust beliefs upon observing a dataset, the Prior-only model should provide the closest match to posterior beliefs.

\subsubsection{Model comparison}

\begin{figure}[t] 
\centering
\includegraphics[width=1\columnwidth]{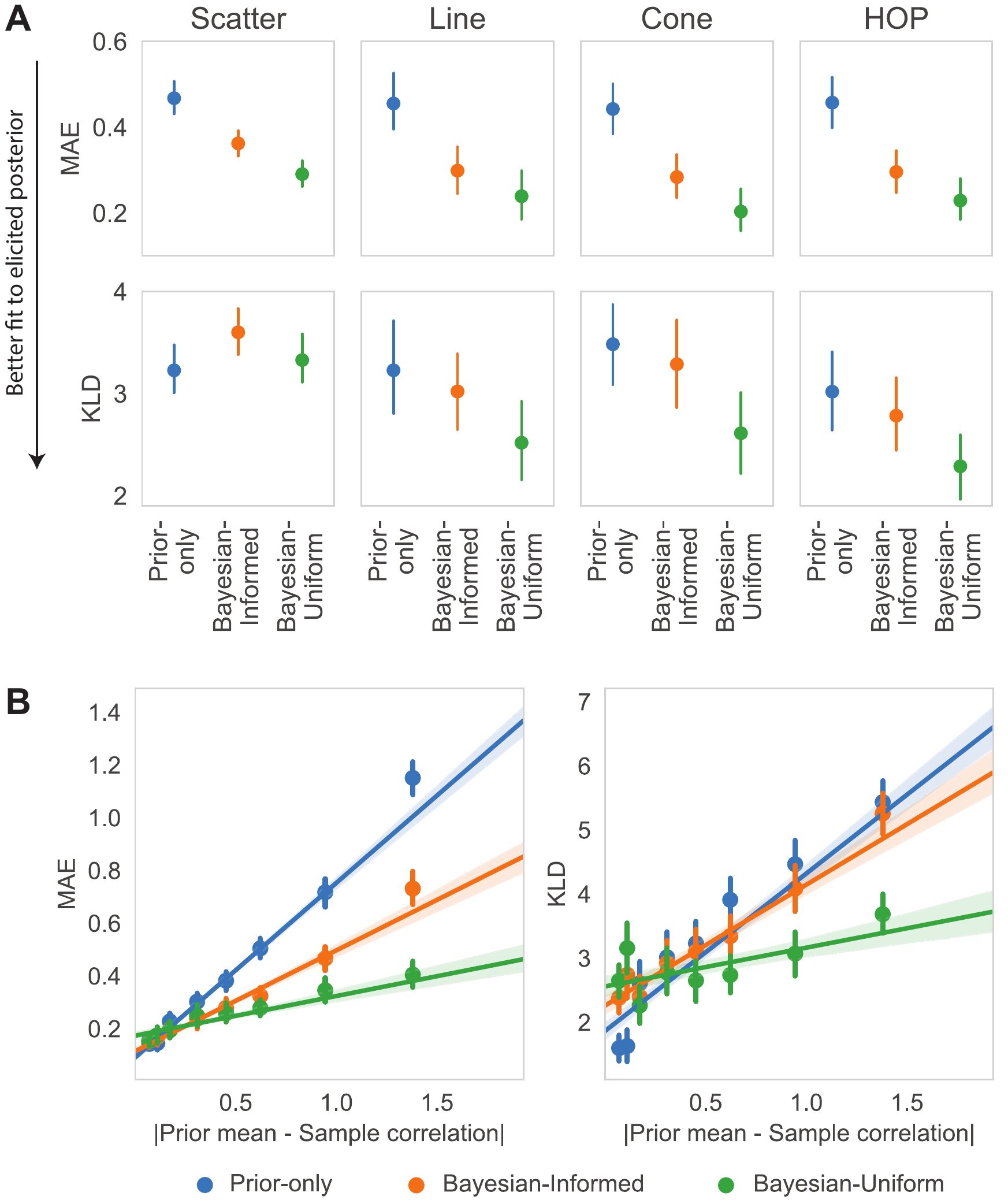}
\caption{\textbf{A:} MAE and KLD between elicited posterior and predictions of Prior-only, Bayesian-Informed, and Bayesian-Uniform models. \textbf{B:} Model performance as a function of the absolute distance between the elicited prior mean and the sample correlation.}
\label{fig:study2_model_results}
\end{figure}

\rev{Following \cite{kim2019bayesian}} we evaluated each model's performance with two metrics: mean absolute error (MAE) between the predicted and elicited posterior means; and Kullback-Liebler distance (KLD) (Figure \ref{fig:study2_model_results}A).
\rev{These measures are complementary in that MAE captures the magnitude of differences in beliefs independently of the amount of uncertainty, while KLD measures correspondence across the entire belief distributions.}
We used mixed effects linear regression to compare MAE and KLD with model and visualization type as fixed effects and random effects for participants and variable sets.

In terms of MAE there were significant effects of visualization treatment ($\chi^2(1,3) = 33.17$, $p < .001$) and model ($\chi^2(1,2) = 333.97$, $p < .001$), but no interaction ($\chi^2(1,6) = 7.55$, $p = .27$).
Pairwise comparisons indicated that MAE was lower under both the Bayesian-Informed and Bayesian-Uniform models than the Prior-only model in all four visualization treatments (all $p < .001$).
The Bayesian-Uniform model achieved lower MAE than the Bayesian-Informed model in the Scatter ($z = -4.52$, $p < .001$), Cone ($z = -2.8$, $p = .01$), and HOP ($z = -2.48$, $p = .04$) conditions, but the two did not differ in the Line condition ($z = -2.22$, $p = .07$).
Comparing the best-fitting Bayesian-Uniform model across visualization treatments showed that MAE was higher in the Scatter group than the Cone ($z = 3.14$, $p = .01$) and Line groups ($z = 2.94$, $p = .02$), but not significantly different from the HOP group ($z = 2.54$, $p = .05$).

For KLD there were significant effects of visualization treatment ($\chi^2(1,3) = 67.57$, $p < .001$), model ($\chi^2(1,2) = 31.64$, $p < .001$), and model $\times$ treatment interaction ($\chi^2(1,6) = 32.49$, $p < .001$).
In the Scatter condition, KLD of the Prior-only model was lower than the Bayesian-Informed model ($z = -3.32$, $p = .003$), but did not differ from the Bayesian-Uniform conditions ($z = -.87$, $p = .66$).
\rev{This indicates that the Bayesian model was relatively unsuccessful at predicting the posterior distribution in the Scatter condition, failing to outperform the baseline Prior-only model.}
In the remaining conditions (Line, Cone, HOP), the Bayesian-Uniform model had lower KLD than both the Prior-only and Bayesian-Informed models (all $p < .022$).
As was the case for MAE, the KLD of the Bayesian-Uniform model was higher in the Scatter condition than the other conditions (all $p < .001$), but did not differ among the Line, Cone, and HOP groups.
This supports the earlier finding that the accuracy of posterior beliefs was poorer in the Scatter condition compared to the other treatments.

The predictions of the three models diverge most when there is a discrepancy between participants' priors and the sample correlation of the observed dataset.
We therefore explored how the fit of each model depended on the absolute distance between the prior mean and the sample correlation (Figure \ref{fig:study2_model_results}B).
At small distances the three models have comparable MAE and KLD, while the advantage for the Bayesian-Uniform model grows with increasing distance between the prior and sample correlation.
The poorer fit of the Bayesian-Informed model indicates that participants discounted their priors when they observed a dataset with a drastically different correlation.
Notably, at small distances KLD was lowest for the Prior-only model.
This result suggests that when people observed a dataset that was consistent with their prior, they were less likely to update their beliefs as predicted by either Bayesian model.


\subsection{Discussion of Study 2}
\rev{Results of the regression analysis and cognitive modeling} showed that visualizations with representations of the population correlation (Line, Cone, and HOPs) led to greater accuracy in posterior beliefs compared to the Scatter condition. 
\rev{In addition, higher correlations led to larger reductions in uncertainty, potentially because stronger relationships are easier to detect in scatterplots \cite{rensink_perception_2010,rensink2017nature} and are associated with less uncertainty in the population correlation.}
We found initial evidence for this updating process using the Bayesian cognitive model, showing that when the sample correlation presented to participants was far from their prior mean, they strongly adjusted their beliefs \rev{to reflect the pattern in the data} (Figure \ref{fig:study2_model_results}B).

\rev{We did not find support for our main hypothesis that uncertainty visualizations would be associated with smaller changes in beliefs.}
On the contrary, the Cone visualization (with a cone of ``plausible alternatives'' representing uncertainty in the correlation based on the data) led to greater reductions in uncertainty. 
\rev{This result suggests that the explicit representation of uncertainty provided by the Cone visualization leads to greater confidence about the true relationship compared to the other visualization types.}
Interestingly, we did not find a similar effect on uncertainty change in the HOP condition, possibly due to the transient nature of the animated uncertainty cone.

\rev{There were two shortcomings of the present study that may have limited our ability to detect differences in belief updating between conditions.}
\rev{First,}
\rev{participant's} prior beliefs 
largely aligned with the sample correlation, leading to many cases with little room for participants to adjust their beliefs.
Second, the relatively large sample size of the datasets ($n = 100$) meant there was relatively little uncertainty about the population correlation.
This may explain why the Bayesian-Uniform model provided the best fit to elicited posteriors, such that the sample correlation had a stronger influence than individuals' priors.
\textbf{Study 3} was designed to further explore how these factors affect belief change. 
We manipulated the data provided to be either congruent or incongruent with the user's elicited prior belief. In addition, we manipulated the amount of data uncertainty by varying the sample size. 

\section{Study 3: How correlation congruence and uncertainty affect belief updating}

The hypothesis of Study 2 was that people would exhibit less belief change when they experienced visualizations with representations of uncertainty. 
The main hypothesis for Study 3\footnote{http://aspredicted.org/blind.php?x=x7ph2u} extends this further to predict that viewers of uncertainty representations would exhibit smaller changes in beliefs when correlation visualizations are \textit{incongruent} with users' prior belief \rev{and when the dataset has a smaller sample size.}

\begin{figure}[t] 
\centering
\includegraphics[width=0.95\columnwidth]{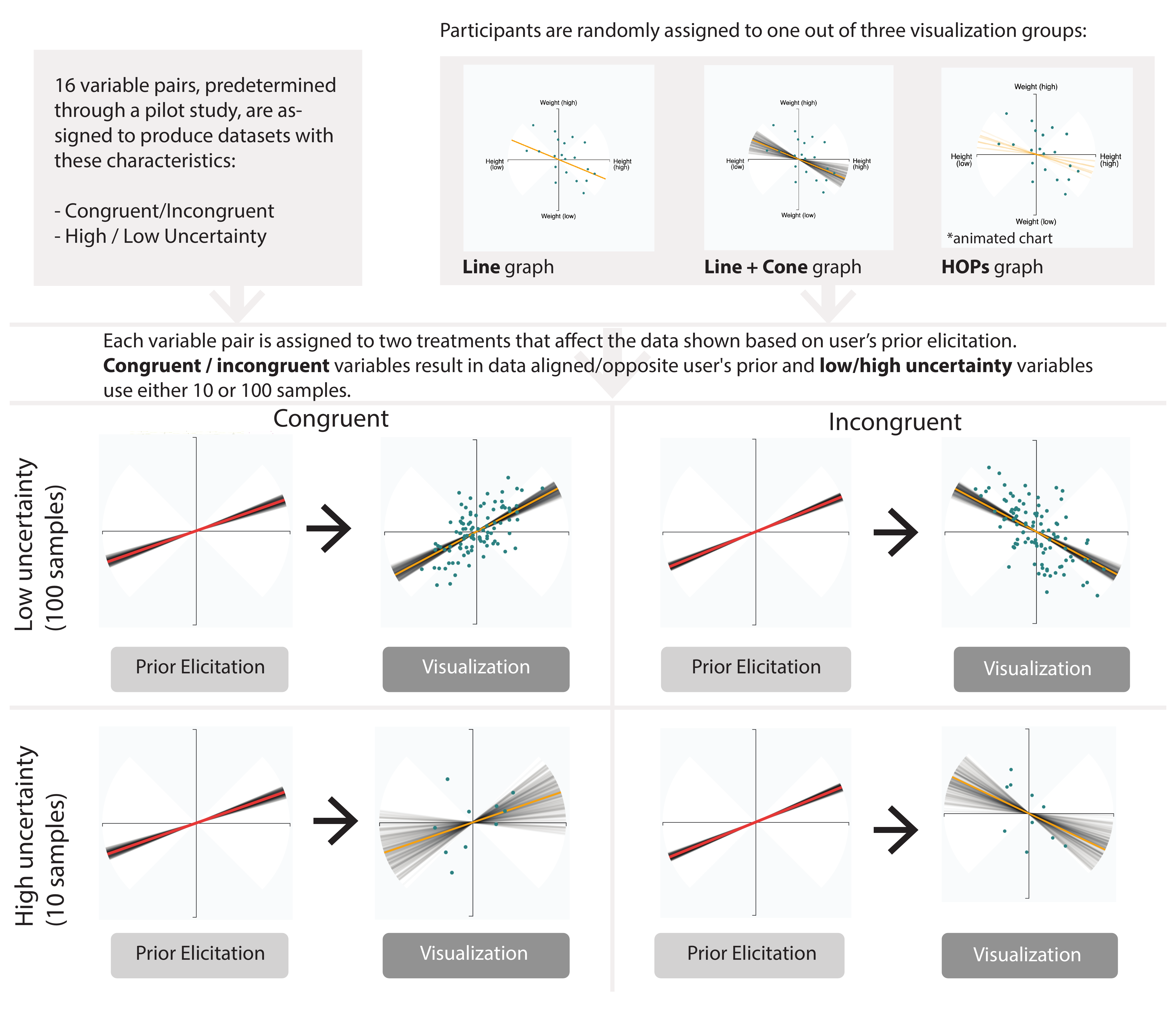}
\caption{Study 3 design. Like Study 2, users elicit their beliefs about correlations of variable pairs before and after seeing data visualizations. Users are randomly assigned to Line, Cone, and HOP visualization treatments. The datasets are generated based on users' prior elicitation as either congruent/incongruent and 10 or 100 data points.}
\label{fig:study3Conditions}
\end{figure}

\subsection{Study Design}
For Study 3, we extended the design of Study 2 by explicitly manipulating the congruence of the sample correlation (\textbf{factor 1}) with a user's prior belief and the amount of uncertainty (\textbf{factor 2}).
Both above factors are within-subjects while the visualization treatment remains a between-subject factor. 
Figure \ref{fig:study3Conditions} summarizes the design of Study 3. 
For each variable pair, participants saw datasets that were either congruent or incongruent to their prior beliefs:
\begin{itemize}
    \item \textbf{Congruent} datasets: Random samples were drawn from a multivariate normal distribution with correlation \textit{0.25} away from the prior mean. For example, if a participant's prior mean was 0.85, the data was sampled from a distribution with population correlation of 0.6 ($0.85-0.25$). In this condition a user always saw sample correlations with the same sign as their prior belief. 
    \item \textbf{Incongruent} datasets: Random samples from a multivariate normal distribution with correlation value that is \textit{1.0} away from the prior mean. For example, if the prior mean was 0.6, the data was sampled from a distribution with a population correlation of \textit{-0.4}. In this condition, participants saw datasets with the opposite correlation sign from their prior belief.
\end{itemize}

We also manipulated the number of samples in the datasets for specific variable pairs (10 points vs. 100 points). 
Datasets with 10 points result in greater uncertainty as measured by the 95\% confidence interval. 
As in Study 2, participants were randomly assigned to visualization conditions of Line, Line + Cone and HOPs. Given Study 2's results that users achieved better accuracy with all three visualization types, we omitted the Scatter condition.

\subsubsection{Datasets, elicitation, and attention check procedures} 
For Study 3 we selected variable pairs from the results of a pilot study.
With 50 pilot participants, we elicited prior belief and uncertainty about 30 variable pair candidates, then categorized variables into a 2 X 2 grid of high/low social consensus on correlation and uncertainty. \footnote{Social consensus was measured as the standard deviation of prior means, while average uncertainty was measured as the mean CI.} 
\rev{With lessons learned on users' beliefs about the variable pairs from Study 2 (section 5.3.1), we aimed to select pairs that cover a range of distributions of beliefs about the mean correlation and uncertainty.} We selected four variables with either high / low correlation consensus and high / low uncertainty. 
Study 3 used the same elicitation process, instructions, and attention checks as Study 2.

\subsection{Participants}
Participants were recruited from Amazon Mechanical Turk.
Participants earned \$1.80 upon completion of the task, which took an average of 22.9 minutes ($SD = 12.28$) to complete.
Per our pre-registration, we excluded any participants who: failed attention check questions ($n=12$); technical errors ($n=95$); or completed the entire task in less than 5 minutes ($n=11$). 
This left $n=267$ participants for the analysis (Line: 89; Cone: 92; HOP: 86).

\subsection{Results}


\setlength{\intextsep}{-4pt}%
\setlength{\columnsep}{-1pt}%
\begin{wrapfigure}{L}{0.16\textwidth}
  \vspace{-11pt}
  \begin{center}
    \includegraphics[width=0.14\textwidth]{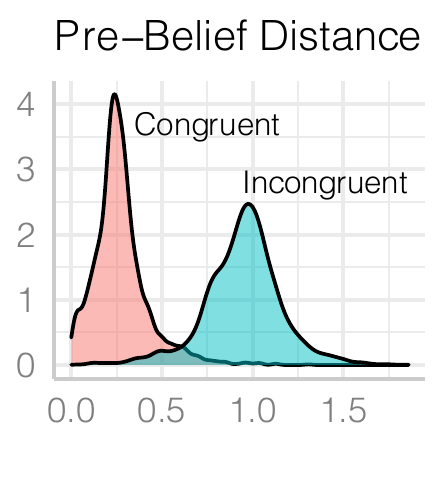}
    \includegraphics[width=0.14\textwidth]{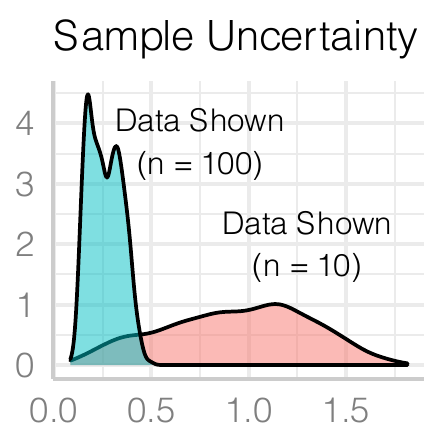}

  \end{center}
\end{wrapfigure}

\textbf{Dependent \& Independent Variables:} Similar to Study 2, we considered three dependent variables: (1) the absolute belief difference, (2) the difference in uncertainty, and (3) the user's belief distance from the model's predicted posterior. For our independent variables (IV), we created two features based on our variable conditions from Figure \ref{fig:study3Conditions}. First, we defined \textbf{pre-belief distance} as the distance between users' prior elicitation and the correlation of the provided sample, which is larger
when a participant is provided incongruent datasets. Next, we defined \textbf{sample uncertainty} as the size of uncertainty shown to users resulting from the sample size. 
In doing so, we used continuous IVs ranging from zero to two rather than binary variables. For reference, we provide kernel density plots (left) for the two IV's partitioned by its respective condition categories.



\textbf{Model Specification:}
We employed three mixed effects models as in Study 2 (see Section 5.3).
For each model, we included the interaction terms between the visualization treatment, the pre-belief distance, and the sample uncertainty as fixed effects. For the visualization treatment, the Line condition is the omitted reference condition. 

\begin{figure} 
\includegraphics[width=1\columnwidth]{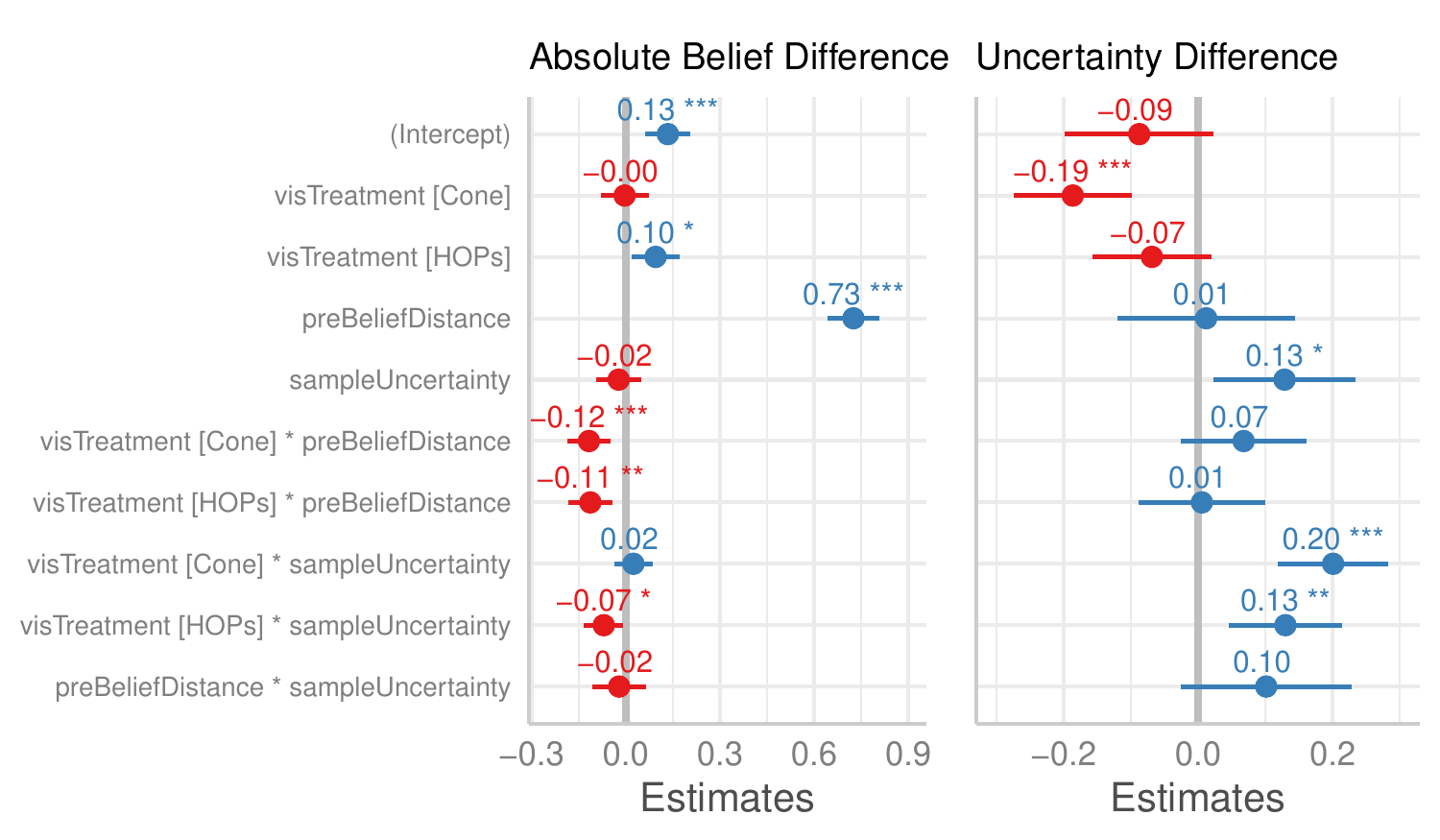}
\caption{Study 3 fixed effects coefficients from analyzing absolute belief difference (left) and uncertainty difference (right). The error bars indicate 95\% confidence intervals. 
Asterisks indicate statistical significance than zero using p-values: *** 99.9\%, ** 99\%, * 95\%. For visTreatment, the reference category is the Line condition.}
\label{fig:study3_me}
\end{figure}

\subsubsection{Change in belief about most likely relationship}


For absolute belief difference (Figure \ref{fig:study3_me}, left), we found the largest effect to be pre-belief distance ($\beta = 0.73 [0.64, 0.81],  z = 2.40, p < .001$), indicating that users updated their beliefs more when they viewed incongruent datasets.

There were significant interactions between pre-belief distance and visualization type, such that there were smaller belief changes when the data was incongruent in both the Cone $(\beta = -0.12 \;\;  [-0.19, -0.05], \;\; z = -3.3, \;\; p < .001)$ and HOPs ($\beta = -0.11 \;\;  [-0.18, -0.04], \;\; z = -3.16, \;\; p < .01$) conditions relative to the Line condition.
This finding is in line with our hypothesis that in the incongruent condition, users would show smaller update in their belief when uncertainty representations are present. 

Finally, while the HOP condition led to slightly larger changes compared to the Line condition $(\beta = 0.10 \;\;  [0.02, 0.17], \;\; z = 2.400, \; p < 0.05)$, this condition had a negative interaction with sample uncertainty \rev{such that beliefs shifted less after seeing smaller datasets} $(\beta = -0.07 \;\;  [-0.13, -0.01], \; z = 2.181, \;\; p < .05)$. 
We did not find corresponding effects for the Cone condition. 
This difference between the Cone and HOP visualizations might suggest that uncertainty is more evident in larger uncertainty amounts when using the HOP technique. This is potentially due to the lack of a fixed representation of most likely correlation in the HOP technique as opposed to the Cone technique.


\subsubsection{Uncertainty change}

In our regression of the uncertainty difference (Figure \ref{fig:study3_me}, right), we found that users in the Cone condition exhibited more reduction in uncertainty than the Line condition $(\beta = -0.19\;\; [-0.27, -0.10], \;\; z = -4.166, \;\; p < .001)$, \rev{replicating the effect seen in Study 2}. There was not a significant effect in the HOPs condition $(\beta = -0.07 \;\; [-0.16, 0.02], \;\; z = -1.514, \;\; p = 0.130)$.

 Pre-belief distance had no effect on the uncertainty difference in any condition. However, sample uncertainty had a positive effect on changes in uncertainty $(\beta = 0.13 \;\; [0.02, 0.23], \;\; z = 2.391, \;\; p < .05)$. We also found that the Cone visualization condition had larger effects on the uncertainty difference when interacting with sample uncertainty $(\beta = 0.20 \;\; [0.12, 0.28], \;\; z = 4.784, \;\; p < .001)$. The HOPs condition also showed a positive interaction with uncertainty difference when interacting with datasets with larger sample uncertainty $(\beta = 0.13 \;\; [0.05, 0.21], \;\; z = 3.019, \;\; p < .01)$. These findings suggest that participants in the Cone condition showed more overall reduction in posterior uncertainty compared to the Line treatment but the HOP condition did not show similar effects. Interestingly, when dealing with larger uncertainty (10 data points), the presence of an uncertainty representation resulted in an increase in users' uncertainty. This finding suggests that both visualization techniques convey uncertainty when uncertainty amounts are larger, but users' experience of the HOP condition is similar to the Line condition when dealing with datasets with smaller uncertainty. Perhaps this is due to users' inability to perceive small angular movements of the line.
 

\subsubsection{Accuracy of posterior beliefs}


We used beta regression to model the effects on the distance of users' posterior beliefs \rev{from the true} sample correlation. 
We found that pre-belief distance had the largest positive effect on users' post-belief distance $(\beta =.34 \;\; [0.16, 0.52] \;\; z= 17.254, \;\; p < .01)$. 
\rev{In other words, posterior beliefs were less similar to the sample correlation when the dataset was incongruent with users' prior beliefs.}
We also found that compared to the Line condition, the HOP condition had a positive effect on posterior distance \rev{when viewing a dataset with more uncertainty} $(\beta = 0.29 \;\; [0.14, 0.45], \;\; z= 2.181, \;\; p < .01)$. This might be due to the lack of a fixed most-likely correlation representation in the HOPs condition, therefore when sample uncertainty is larger, users are more prone to larger distances (errors) in their judgements.



\subsection{Bayesian belief updating model}

\begin{figure} 
\centering
\includegraphics[width=0.85\columnwidth]{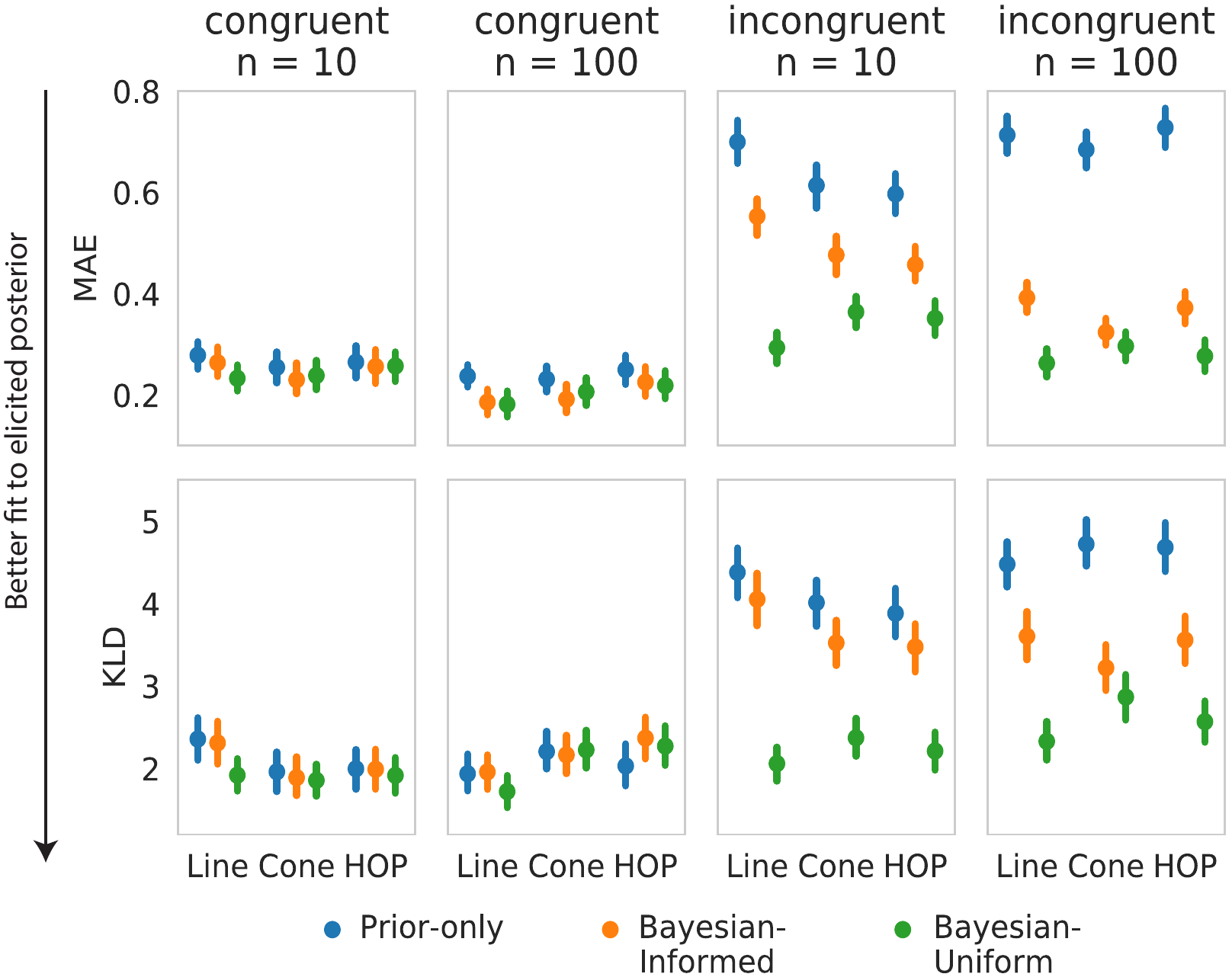}
\caption{MAE and KLD by model for Study 3.}
\label{fig:study3_models}
\end{figure}

We used the models from Study 2 to examine how prior beliefs influenced belief updating in Study 3. 
In general, the best fit to elicited posteriors in terms of both MAE and KLD was achieved by the Bayesian-Uniform model in all conditions (Figure \ref{fig:study3_models}). 
Incongruent trials provide a strong comparison of the Bayesian-Informed and Bayesian-Uniform models because they involve datasets that conflict with participants' prior beliefs.
If people integrate new evidence with their elicited prior, they should show smaller shifts in beliefs in Incongruent trials than expected under the Bayesian-Uniform model.
However, as was seen in Study 2, posterior distributions were best-fit by the Bayesian-Uniform model, suggesting a stronger influence of the data visualization on posterior beliefs. 
Notably, the only condition in which the two models performed comparably on Incongruent trials was the Cone treatment, where there were no differences in MAE ($t(722.63) = 1.33$, $p = 0.18$) or KLD
($t(731.98) = 1.78$, $p = 0.08$), indicating that Cone visualizations produced belief updates that more closely aligned with the normative prediction of the Bayesian-Informed model.


\subsection{Discussion of Study 3}

We predicted that people exposed to uncertainty visualizations (Cone and HOP conditions) would exhibit less belief change compared to those without uncertainty (Scatter and Line conditions).
We found strong support for this hypothesis in Study 3 when participants saw data that was incongruent with their prior beliefs. 
Both the Cone and HOP treatments were associated with smaller belief updates compared to the Line condition which did not represent uncertainty about the correlation.
Uncertainty visualizations also affected whether there were shifts in participants' degree of uncertainty.
Relative to the Line condition, Cone visualizations led to greater reductions in uncertainty for large datasets, whereas uncertainty did not change when datasets were small. 
Similar (albeit weaker) effects were present for HOP visualizations. 

Finally, we replicated the modeling results from Study 2, showing that posterior beliefs were best-fit by the predictions of the Bayesian-Uniform model.
Although this does not imply that participants completely disregarded their prior beliefs, it indicates that the data visualizations tended to have a stronger influence on posterior beliefs than expected from a normative Bayesian perspective.
\rev{The Cone visualization was the only condition in which the Bayesian-Informed model performed comparably to the Bayesian-Uniform model.
This result suggests an alternative interpretation of the smaller degree of belief updating in that condition when faced with incongruent data. Rather than representing an irrational failure to modify beliefs akin to confirmation bias, the Cone condition may be most effective for striking the appropriate balance between new data and prior beliefs.}

\section{Discussion, Future Work, and Conclusion}

In this paper, we study the effect of prior belief and uncertainty representations on correlation judgement.
In Study 1 we developed the Line + Cone method for eliciting people's beliefs about the correlation between two variables, including their degree of uncertainty. \rev{The Line + Cone method serves as a good choice for eliciting users' beliefs about bivariate relationships for future studies of correlation judgement. In addition to capturing users' beliefs about the correlation means (commonly done in previous correlation judgement studies), results from all three studies demonstrate that it is also important to capture users' uncertainties about their judgements.}
In Studies 2 and 3, we used the Line + Cone method to investigate belief updating in the context of data visualization. 
We found that visualization conditions with uncertainty communication led to less belief updating compared to visualizations without uncertainty, especially when the presented correlation visualization is incongruent with users' prior beliefs. \rev{An important conclusion is that judgements are affected by the existence of uncertainty depictions. How we encode uncertainty (e.g., Cone vs. HOPs), also affects users' belief and uncertainty change. As the visualization community pays more attention to the importance of uncertainty representations and elicitation, it is important to be cognizant to the affects of such techniques on users' judgements.}

In our studies we applied a Bayesian cognition framework to understand how people update their beliefs about bivariate correlations with different types of visualizations.
Recent studies have applied insights from Bayesian cognitive modeling to understand how people integrate new data with their existing knowledge \cite{griffiths2008bayesian,lieder2018resource}.
The Bayesian framework provides normative benchmarks that can be used to evaluate whether people optimally revise their beliefs given their existing uncertainty and the strength of new evidence conveyed through a visualization \cite{kim2019bayesian}.
We used Bayesian models to compare participants' posteriors to three benchmarks: no change in beliefs (Prior-only model); the normative posterior when taking into account the elicited prior (Bayesian-Informed model); and the normative posterior when disregarding the prior (Bayesian-Uniform model).
In both Studies 2 and 3, elicited posterior distributions were best-described by the Bayesian-Uniform model, suggesting that the characteristics of the visualized dataset had a stronger influence on posterior beliefs than expected under the Bayesian-Informed model.

There are several possible explanations for why posterior beliefs appeared to underweight participants' priors.
One possibility is that people have a different interpretation of the cone representation which is used to elicit their uncertainty. 
In order to minimize demands on numerical or probabilistic reasoning, participants were simply instructed to adjust the cone to capture the range of ``plausible alternatives'' for the correlation between the variables. 
In Study 1 we found support for the claim that this method captures participants' uncertainty, but there may nevertheless be a mismatch between the elicited distribution and participants' subjective beliefs such that people are more uncertain than indicated by their elicited priors.

We found other evidence that people updated beliefs in a way consistent with Bayesian inference.
In Study 2, users reduced their uncertainty to a greater extent for more extreme sample correlations.
In Study 3, uncertainty increased when people saw small datasets ($n = 10$) compared to large datasets ($n = 100$), even in the Line condition which lacked an explicit representation of the correlation uncertainty.
Participants also expressed greater uncertainty in the posterior beliefs than predicted by the Bayesian models, echoing the findings of Kim \textit{et al.} \cite{kim2019bayesian}. 

These studies provide the groundwork for investigating how people interpret data that is relevant to strongly-held or favored beliefs.
Prior beliefs can distort the perception of new evidence, as is seen in widespread evidence of confirmation bias \cite{mynatt1977confirmation,nickerson1998confirmation,heuer1999psychology}. 
Using intuitive, visual belief elicitation methods in conjunction with Bayesian cognitive models offer a promising path toward understanding the causes of such biases in data visualization.

\bibliographystyle{abbrv-doi}

\bibliography{template}
\end{document}